\documentclass{emulateapj}

\usepackage{graphicx}
\usepackage[hypertex]{hyperref}

\def \th {\thinspace}
\def \nh {N${\rm _H}$}

\def \arcmin {\hbox{$^\prime$}}
\def \arcsec {\hbox{$^{\prime\prime}$}}

\def\spose#1{\hbox to 0pt{#1\hss}}
\def\ltsim{$\mathrel{\spose{\lower 3pt\hbox{$\sim$}}
        \raise 2.0pt\hbox{$<$}}$\thinspace}
\def\gtsim{$\mathrel{\spose{\lower 3pt\hbox{$\sim$}}
        \raise 2.0pt\hbox{$>$}}$\thinspace}

\def \nh {$N_{\rm H}$}
\def \src {NGC\th 1332}
\def \srcthree {NGC\th 720}

\def \eg {e.g.}

\def \ie {i.e.}
\def \cf {c.f.}
\def \dtwentyfive {${\rm D_{25}}$}
\newcommand{\zfe }{${\rm Z_{Fe}}$}
\newcommand{\zo }{${\rm Z_{O}}$}
\newcommand{\zsi }{${\rm Z_{Si}}$}
\newcommand{\zne }{${\rm Z_{Ne}}$}
\newcommand{\zmg }{${\rm Z_{Mg}}$}
\newcommand{\chandra }{{\em Chandra}}
\newcommand{\xspec }{{\em Xspec}}
\newcommand{\acis }{{\em ACIS}}
\newcommand{\ciao }{{\em CIAO}}
\newcommand{\caldb }{{\em Caldb}}

\newcommand{\xmm }{{\em XMM}}
\newcommand{\asca }{{\em ASCA}}
\newcommand{\rosat }{{\em Rosat}}

\newcommand{\lx }{${\rm L_X}$}
\newcommand{\lb }{${\rm L_B}$}

\newcommand{\ned}{{\em{NED}}}

\shorttitle{Solar abundances in \src}
\shortauthors{Humphrey et al.}

\begin{document}

\title{A Chandra view of the normal S0 galaxy NGC\th 1332: II: Solar abundances in the hot gas and implications for SN enrichment.}
\author {Philip J. Humphrey\altaffilmark{1},  David A. Buote\altaffilmark{1} and Claude R. Canizares\altaffilmark{2}}
\altaffiltext{1}{Department of Physics and Astronomy, University of California at Irvine, 4129 Frederick Reines Hall, Irvine, CA 92697-4575}
\altaffiltext{2}{Department of Physics and Center for Space Research, 37-241
Massachusetts Institute of Technology, 77 Massachusetts Avenue, Cambridge,
MA 02139}
\begin{abstract}
Using a new Chandra \acis-S3 observation of the normal, isolated, 
moderate-\lx,
lenticular galaxy \src\, we  resolve the emission into
$\sim$75 point-sources, and a significant diffuse component. 
We present a detailed analysis of the spectral properties of the diffuse
emission, constraining both the temperature profile and the metal abundances
in the hot gas. 
The characteristics of the point source population and the 
spatial properties of the diffuse emission are discussed in two companion 
papers.
The diffuse component comprises hot gas, with an $\sim$isothermal temperature 
profile ($\sim$0.5~keV), 
and emission from unresolved point-sources. In contrast with the 
cool cores of many groups and clusters, we find a small central temperature
peak. We obtain emission-weighted abundance contraints within 20~kpc for
several key elements: Fe, O, Ne, Mg and Si. The measured 
iron abundance (\zfe=1.1 in solar units; $>$0.53 at 99\% confidence) 
strongly excludes the very sub-solar values often historically 
reported for early-type galaxies.  
This continues, in a lower-\lx\ system,
a trend in recent observations of bright galaxies and groups. 
The abundance ratios, with respect to Fe, of the other elements were
also found to be $\sim$solar, with the exception of \zo/\zfe\ which was
significantly lower ($<0.4$), as seen in several bright galaxies, 
groups and clusters. Such a low O abundance is not predicted by simple
models of ISM enrichment by Type Ia and Type II supernovae,
and may indicate a significant contribution from primordial hypernovae. 
Revisiting \chandra\ observations of the moderate-\lx, isolated elliptical
\srcthree, we obtain similar abundance constraints 
(\zfe=$0.71^{+0.40}_{-0.21}$, 90\% confidence; \zo/\zfe=$0.23\pm0.21$).
Adopting standard SNIa and SNII metal yield models, our abundance ratio
constraints imply 73$\pm5$\% and 85$\pm$6\% of the Fe enrichment in \src\ and 
\srcthree, respectively, arises from SNIa. Although these results are 
sensitive to the considerable systematic  uncertainty in the SNe yields, 
they are in good agreement with observations of more massive systems.
These two cases of moderate-\lx\ early-type galaxies reveal a consistent
pattern of metal enrichment from cluster scales to moderate \lx/\lb\ galaxies.
\end{abstract}


\section{Introduction}
During the formation and evolution of early-type galaxies, the primordial gas 
is enriched by  supernovae ejecta and stellar mass-loss.
Observations of the metal content of the hot, X-ray emitting gas within
such galaxies therefore offer a powerful means to investigate their history
\citep[\eg][]{loewenstein91}.
In early \asca\ and \rosat\ observations  of early-type galaxies
very sub-solar metal abundances (dominated by Fe, which has the strongest
diagnostic lines in the soft X-ray band) were generally reported
\citep[\eg][]{davis96b,mulchaey98,loewenstein98,matsumoto97,matsushita94}.
In stark  contrast, the mean stellar iron abundance in elliptical
galaxies tends to be $\sim$solar  \citep{arimoto97},
implying that the ISM could not have been substantially 
enriched by SNIa and mass-loss from the stellar population.
However, the mean iron abundances determined by 
X-ray observations of clusters tend to be $\sim$0.3--0.5 times solar, 
requiring significant enrichment of the primordial gas, most probably  
from the stellar population of the 
giant elliptical galaxies \citep{renzini97}. 
This discrepancy is problematical, as it is difficult
to envisage how individual early-type galaxies could have, 
in many cases, {\em lower} iron abundances than typical clusters. Indeed,
gas enrichment models for the centres of individual galaxies tend to predict
super-solar values of \zfe\ \citep[\eg][]{ciotti91,loewenstein91,brighenti99a}.
This obvious inconsistency led \citet{arimoto97} to question the 
validity of the X-ray plasma codes which were being fitted to the data,
especially given uncertainties in the physics of the Fe L-shell transitions.

An alternative explanation was provided by \citet{buote98c}, who
demonstrated that fitting a single temperature model to intrinsically
non-isothermal data can give rise to an ``Fe bias'', in which the 
iron abundance is systematically under-estimated, and instead obtained
$\sim$solar abundances by fitting multi-temperature models to a variety of 
bright elliptical galaxies (see also \citealt{buote00a} and \eg\
\citealt{buote03b}, who found comparable results with different plasma models).
Although it had long been recognized that a hard spectral component 
(to account for unresolved point source emission) was 
required to fit the \asca\ data of many early-type galaxies 
\citep{matsushita94}, the spectral shape of this component is much
harder (kT\gtsim 5~keV) than the hot gas emission 
(for which kT$\sim$0.5--1.0~keV),
so that it cannot move to mitigate, even in part, the Fe bias.
Nonetheless, \citet{buote98c} still found that, in the lower-\lx\ systems,
in which the data were
consistent with a {\em single} hot gas component plus unresolved
sources, the abundances were consistent with \zfe$\simeq$1. However,
the lower signal-to-noise resulted in very poor constraints so that 
very sub-solar \zfe\ values could often not be entirely ruled out.
A similar result for a low S/N system was reported  by \cite{kim96} for the  
 low \lx/\lb\ S0 NGC\th 4382 observed with \asca.  

More recent \chandra\ and \xmm\ observations of bright ellipticals
and the centres of bright groups have tended to confirm near solar 
(or even super-solar) abundances 
(\eg\ NGC\th 5044:\citealt{buote03b},\citealt{tamura03a}; 
M\th 87: \citealt{gastaldello02a}; NGC\th 1399: \citealt{buote02a}; 
NGC\th 4636: \citealt{xu02a}; MKW\th 4: \citealt{osullivan03b};
also \cf\ the low \lx/\lb\ radio galaxy NGC\th 1316: \citealt{kim03b}). 
However, for some systems, especially (but not exclusively) the 
lowest-\lx/\lb\ galaxies, authors are 
still tending to find very sub-solar metal abundances
(\eg\ NGC\th 1291: \citealt{irwin02a},
NGC\th 4697: \citealt{sarazin01}; especially NGC\th 3585,
NGC\th 4494 and NGC\th 5322, for which \zfe\ltsim 0.1 was 
reported by \citealt{osullivan04a}; also \cf\ the X-ray bright radio
galaxy NGC\th 6251: \citealt{sambruna04a}).

The lack of a consistent picture of enrichment from galaxy
to cluster scales is a major problem in our understanding 
of galaxy evolution. 
It is therefore of critical importance to determine accurate 
abundances in galaxies with a wide range of \lx/\lb, and especially
to determine the ubiquity of the very sub-solar abundances in 
low-\lx/\lb\ systems. 
In this paper, we present a \chandra\ study of the metal abundances in the
hot gas of the ``normal'', relatively isolated, 
moderate-\lx\ lenticular galaxy \src. 
In two companion papers 
\citep[][hereafter Paper I and Paper III]{humphrey04a,buote04a},
we present a study of the X-ray point-source population and the 
gravitating matter distribution, which have been able to take advantage of the 
excellent spatial resolution of \chandra. 

\src\ has a much lower
\lx\ (and \lx/\lb) than most of the gas-rich giant elliptical galaxies
for which $\sim$solar abundances have been found. Furthermore, its 
\lx/\lb\ is sufficiently high to ensure substantial emission from the hot
gas. For many systems with much lower \lx/\lb, such analysis is hampered
by  a lack of photons \citep[which may introduce abundance degeneracies or 
extremely poor constraints; \cf][]{buote98c}.
Significant diffuse gas emission in \src\ 
was inferred from previous X-ray 
observations, making it an ideal subject for such a study. 
Using \rosat\ PSPC observations, \citet{buote96a} found evidence 
of a flattened dark matter halo in \src.
Although the \asca\ data enabled the unambiguous identification
of a hard spectral component (supposed to arise from unresolved point
sources), the signal-to-noise was insufficient to enable good constraints
on the metal abundances to be determined 
\citep{buote98c} (rescaling to the new solar abundances,
these authors found \zfe$>0.3$). Although the
best-fit value (\zfe$\simeq$1.2) was clearly consistent with $\sim$solar
abundances, highly subsolar values could not be excluded.

Throughout this paper, we adopt the new standard abundances
of \citet{grsa}. We note that there is some complication in comparing to
the literature as many authors still use the outdated solar
abundances of \citet{anders89}. Although we quote all abundances 
with respect to the new standard, the most significant discrepancy between
fits using  the two standards is a difference in the inferred value of \zfe; 
for comparison with our work,
\zfe\ determined using the older abundances should be multiplied by 
$\sim$1.5. All quoted confidence regions are 90\%, unless otherwise
stated.

\section{Background and Flaring} \label{sect_flare}

\begin{figure*}
\plottwo{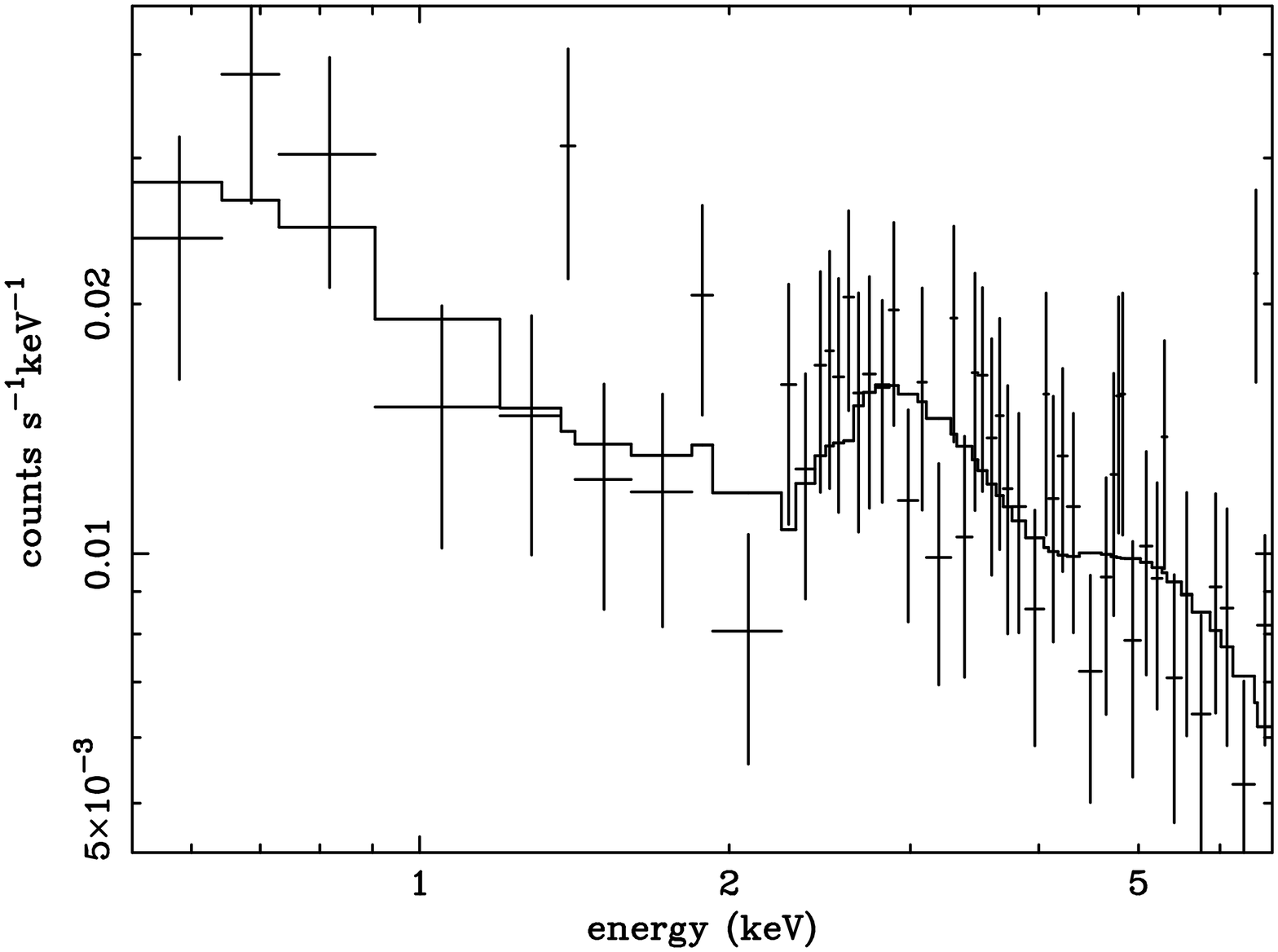}{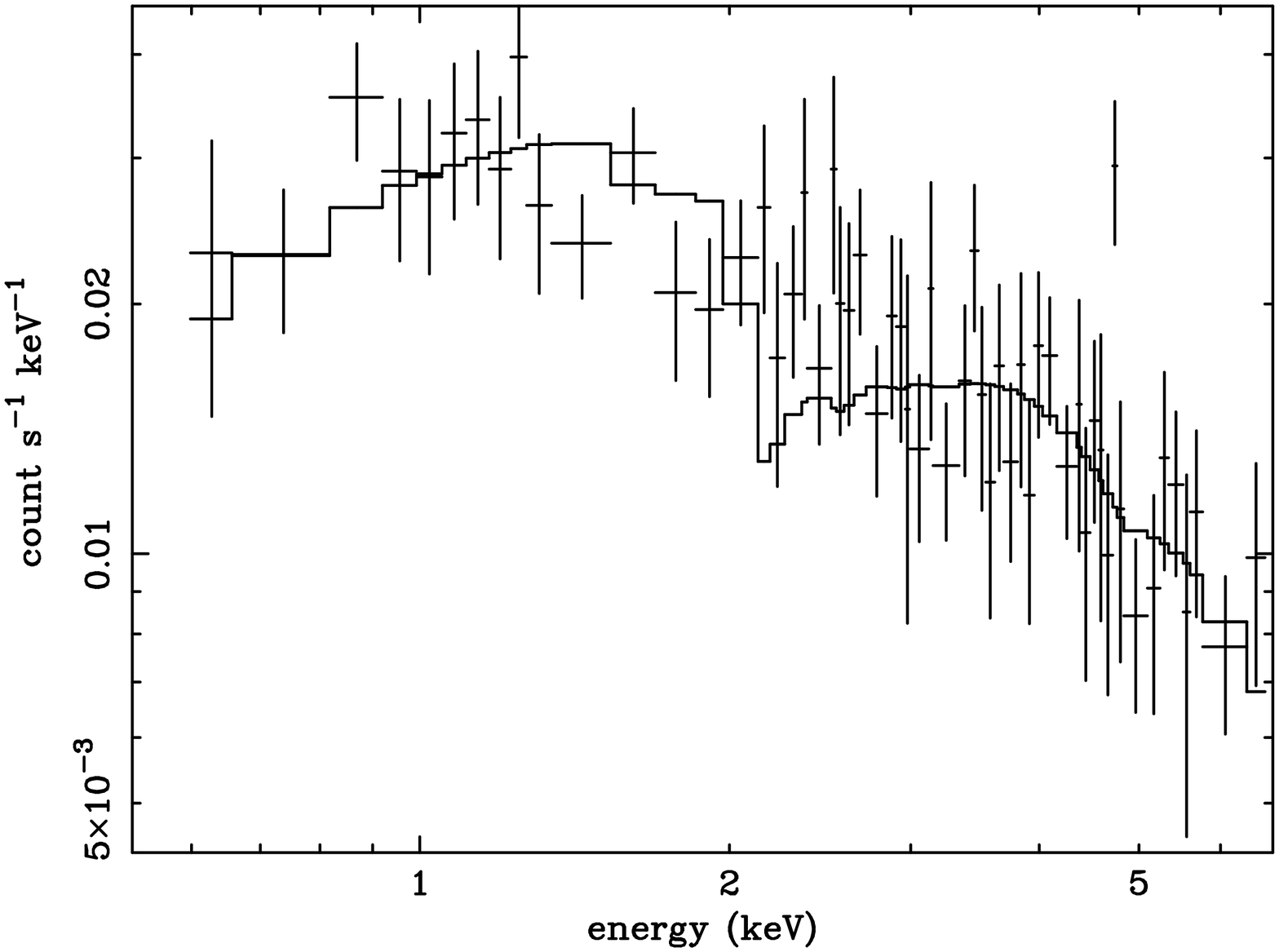}
\caption{The spectra of the flaring, 
shown along with the best-fit models.  Left panel: the flare
spectrum obtained from the S3 chip. Right panel, the flare spectrum
obtained from the S1 chip. Note that, since the extraction regions were
not identical, we do not expect the spectral normalizations 
to be identical.} \label{fig_flare}
\end{figure*}

The region of sky containing \src\ was observed with the ACIS instrument
aboard \chandra\ between  2002 September 19 10:39 and September 20 02:59 UTC,
the galaxy being centred on the S3 chip, for a nominal $\sim$60~ks exposure.
An additional observation made between September 18 02:56 and 08:27 UTC,
for a nominal 20~ks exposure, was heavily contaminated by flaring of the 
background and so we present results only for the better-quality dataset.
We reduced and processed the data using the \ciao\ data-analysis suite of 
tools,
version 3.0.2, and the \chandra\ \caldb\ version 2.26, and we performed
spectral-fitting with XSPEC version 11.3.0. 
For a discussion of the  data-reduction procedures and a detailed analysis 
of the point-source population in \src, we refer the interested
reader to Paper I. 

One of the challenges in performing analysis of relatively 
low surface-brightness diffuse emission is the determination of a reliable
background estimate. Since the diffuse emission covers much of the S3 chip,
it was not possible to determine a ``local'' background estimate from 
an appropriate region of the detector. We therefore adopted the 
appropriate ``blank field'' event files provided in the \caldb.
There is known to be variation between different observations and 
different pointings
in both the high and low-energy regimes of the background.
This is complicated further by the ``flaring'' which is
often observed in the background \citep[\eg][]{markevitch02}. 
We therefore accumulated a background lightcurve (from 5.0--10.0~keV) 
from source-free regions of the active chips and examined it by eye for 
evidence of flaring. The lightcurve shows evidence of mild flaring activity.
The strongest flares had amplitudes reaching $\sim$50\%, but lasted
for only a short fraction of the entire observation. In contrast, low-level
flaring persisted through much of the observation, with an associated
amplitude of $\sim$20\%. 

Completely excising the low-level flaring from the data resulted in 
a total exposure of only $\sim$28~ks. We found excellent agreement in the
high-energy count-rates of the background templates and the flare-cleaned
observation, indicating complete removal of the flare. However,
such a short exposure unacceptably degraded the signal-to-noise level 
of the data. As a compromise, we excised only the periods of strongest 
flaring, which  resulted in a more acceptable $\sim$45~ks exposure.
{We chose not to include all of the data since the significantly higher background during the remaining time periods would more than offset the signal-to-noise gain from the modest increase in exposure that would result}.
However, we compared spectra accumulated with each level of filtering
from the same region of the S3 chip containing low surface-brightness
emission from \src. The less-filtered data clearly showed significant
residuals in excess of the hot gas emission at \gtsim 2~keV which
were not seen in the more heavily-filtered data. 
In order to take account of this contamination, we chose to include
an additional component in our spectral-modelling. We did not
attempt to renormalize the background template to match the high-energy
residuals as this will also renormalize the soft X-ray background and 
instrumental features (which can be problematic if the flaring 
is sufficiently strong, as in Sect~\ref{sect720}).

To enable us to estimate the spectral form of the flaring, it was 
necessary to disentangle it from the diffuse galactic emission and the 
X-ray background. To achieve this, we adopted two complementary 
techniques. Firstly, we exploited the strong variability of the 
flaring component, in contrast to (naturally) constant galactic 
emission. 
We accumulated a spectrum at the peak of flaring and one during a period
when flaring was absent, from the same annular region from 5--160\arcsec\
(excluding all photons within a region
6 times the radius of the 1$\sigma$ detection region around
each resolved point-source, since individual sources may be 
variable). The difference between these two spectra must, necessarily,
be the spectrum of the flare.
The two periods were chosen from the lightcurve
by eye (we note that intensity-filters would not have been appropriate
since the Poisson scatter, for realistic lightcurve binsizes, was not
much less than the flaring amplitude). Appropriate count-weighted 
spectral response matrices were generated for the peak flaring data 
with the \ciao\ tasks {\bf mkrmf} and {\bf mkwarf}, and the flare-free
spectrum was adopted as the background. 
The ``flare'' data were regrouped to ensure at least 20 
source-plus-background photons in each
spectral bin, and a signal-to-noise ratio of at least 3. 
{This procedure assumes that the time-averaged spectrum of the flare is similar to the peak flare spectrum. A comparison of flare lightcurves accumulated in different energy bands (0.5--3.0~keV and 3.0--10.0~keV), obtained by subtracting the ``quiescent'' level (estimated from non-flaring stretches of data) from the background lightcurves, did not reveal any significant changes in spectral hardness during evolution of the flare. This would seem to support our assumption.}

To parameterize the spectral shape of the 
flare we tested various spectral models against the data (which
were not required to be physical). We found that the data could be very 
well-fitted ($\chi^2$/dof = 41.5/46)  with a broken-power law model,
plus a broad gaussian term at  $\sim$2.6~keV to  soften the change in the 
spectral slope. 
In order to determine whether the flare spectrum varied  with
position on the S3 chip, we extracted similar spectra from
the set of 5 annuli adopted in Sect~\ref{sect_spectra}. Excluding the 
two innermost annuli (for which the data were too poor for comparison),
we found that the parameterized model gave an excellent fit to the data
within each annulus, if the overall normalization was allowed
 to fit freely. The normalization scaled approximately with the 
extraction area, as expected for flaring which was uniform 
across the S3 chip.

Our alternative means of estimating the flare spectrum was to accumulate
the spectrum from a source-free region of the S1 chip (since this was 
sufficiently offset from the galaxy centroid not to be strongly 
contaminated by emission from the hot gas) and account for the 
X-ray background by adopting the appropriate background template. 
We accumulated a spectrum
from a region centred on the S1 chip, with a 160\arcsec\ radius.
We  excluded data in the vicinity of  a single point-source found 
by the detection algorithm on this chip. We extracted a background
spectrum from the standard S1 ``blank-fields'' events file. 
The source spectrum was grouped to ensure there were at least 20 photons in 
each data bin and the signal-to-noise ratio exceeded 3. 
We were able to parameterize  the S1 flare spectrum with a 
single broken power law model.  There was excellent
agreement between flare spectra accumulated {\em via} both methods above
$\sim$2~keV (Fig~\ref{fig_flare}). Between $\sim$1--2~keV, the flare accumulated from the 
S1 chip showed a significantly higher count-rate than that obtained from
the S3 chip. A number of systematic effects may account for this
discrepancy. Firstly, the S1 chip may not be entirely free of contamination
by the diffuse emission from the galaxy. Secondly, the degradation of the 
PSF on the S1 due to the large offset means that point-sources may not
have been detected on this chip, contaminating the data. Thirdly, the actual
flare spectrum may be different between the two chips. 
Finally, there may be subtle differences between the
background template and the actual background spectrum for our pointing. 
In the subsequent sections, we adopt the flaring model taken from the 
S3 chip, since, on balance, it is less likely to be subject to biases
than the data from the S1 chip. However, we found very little 
difference if the alternative flare model was used 
(Sect~\ref{sect_flare_bg}). {Given the very different assumptions involved in our two estimates of the flare spectrum, this gave us confidence in our ability to model the flaring}.

\section{Spectral analysis} \label{sect_spectra}

\begin{figure*}
\plottwo{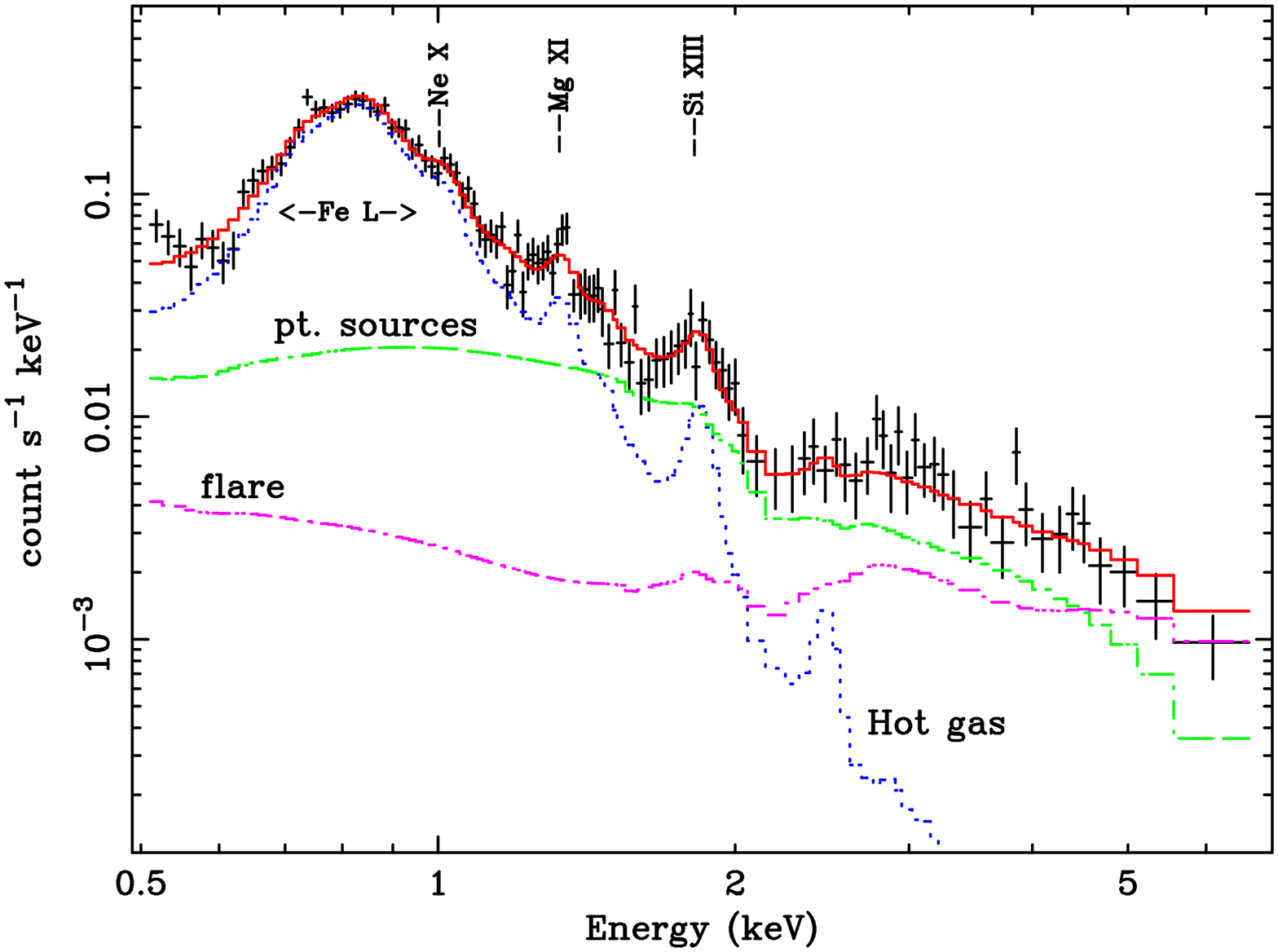}{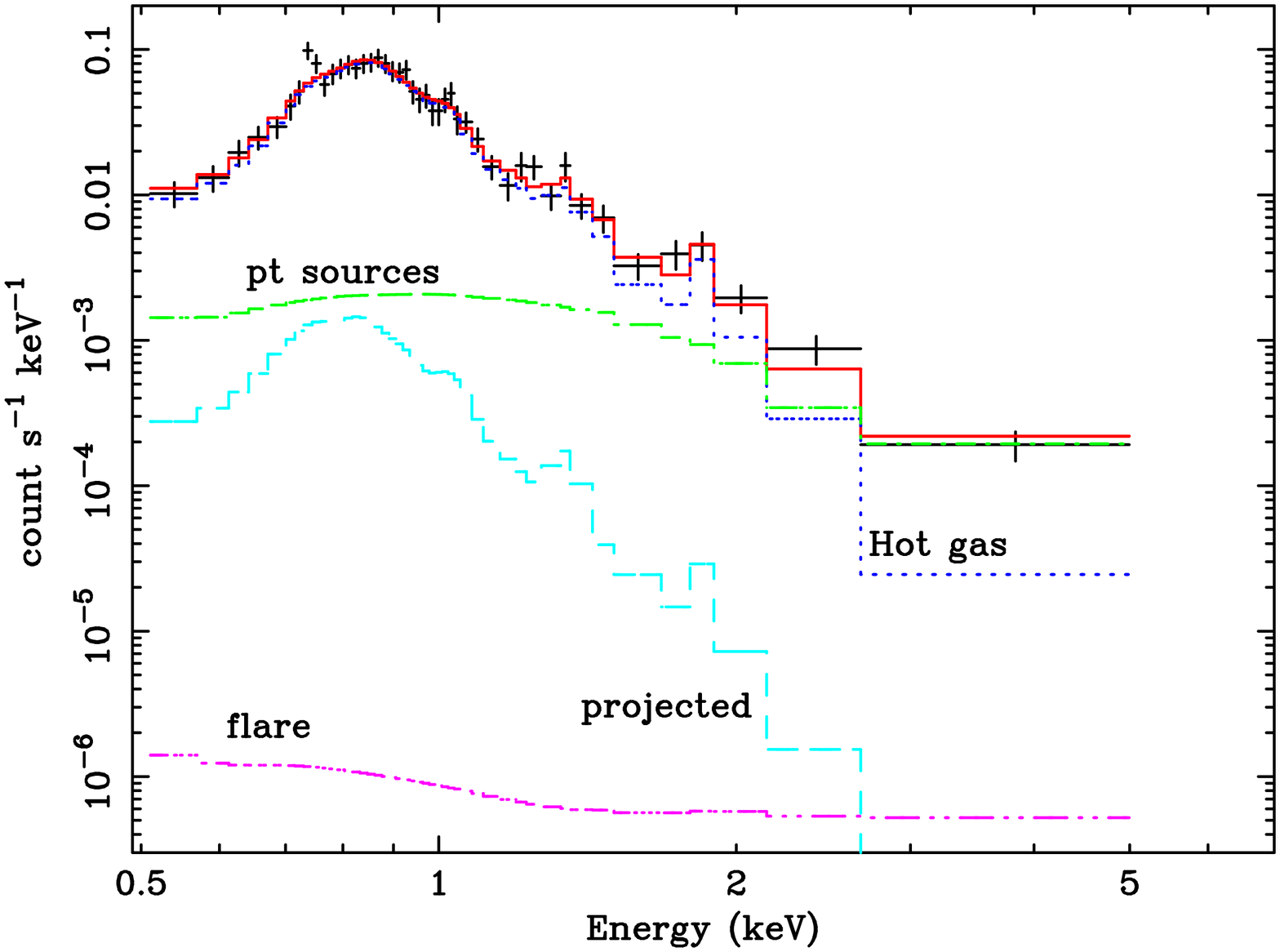}
\caption{Left panel: spectrum of the large aperture. Right panel: 
spectrum of the innermost annulus. Also shown are the best-fit
spectral-models (solid line; we use the single-temperature fit for the large 
aperture) and the individual contributions due to the hot gas
(dotted line), the ``contamination'' due to projection effects (dashed
line, right panel only), the unresolved point-sources (dash-dot line) and flaring
(dash-dot-dot). The significance of the flaring component rises in the
outermost annuli. 
Note that the feature at $\sim$0.72 keV in the 
left panel is too narrow to be an emission  line; it is a statistical fluctuation at 
$<3\sigma$ in a single data-bin. Since these spectra are not independent,
it is visible in both panels, but it is not seen in other annuli.
}\label{fig_spectra}
\end{figure*}

For analysis of the diffuse galactic emission of \src, 
we extracted spectra from a single, large (3\arcmin) aperture and 
a series of contiguous, concentric annuli, all centred on the centroid of the 
X-ray emission. We determined the galaxy centroid 
self-consistently by iteratively computing the centroid in a 1\arcmin\ circle,
which was re-centred at this new position, until no more refinements were
required. Our initial guess
for the centroid of X-ray emission was the 2MASS position given in \ned.
Our resulting position was in excellent agreement with that of the
diffuse source identified at the galaxy centre by the point-source 
detection algorithm (Paper I) and was within $1$\arcsec\ of the infra-red 
centroid. In order to minimize contamination from point sources we excluded 
regions around each detected point source (Paper I) of radius 6 times 
the semimajor axis of the 1-$\sigma$ encircled energy ellipse.
The size of the large aperture was chosen to maximize signal-to-noise,
whilst the individual annuli were chosen so as to contain in each 
approximately $\sim$1500 background-subtracted photons. This gave
annuli with outer radii 0.07\arcmin, 0.55\arcmin, 1.4\arcmin, 2.3\arcmin\
and 3.1\arcmin. 
Appropriate count-weighted response matrices were generated for each
spectrum, using the standard \ciao\ tools {\bf mkrmf} and {\bf mkwarf}. 
The spectra within each annulus
were rebinned to ensure at least 20 photons in each spectral bin and
to ensure a signal-to-noise ratio of at least 3. 

We restricted our fitting to the energy-band 0.5--7.0~keV, on 
account of calibration uncertainties at lower energies and the 
steeply-rising background at high energies. 
At low energies, the background template spectrum becomes less reliable
since there is known to be strong dependence in the soft X-ray background 
emission with pointing. We therefore took care to ensure that the data
were truncated at low energies where the background and source began to
converge. For the spectra of the two outermost annuli, 
this involved truncating the data
below $\sim$0.7~keV; for the innermost annuli data were excluded below
0.5~keV. Care should be exercised when comparing results obtained for 
different energy bandwidths, as narrowing the bandwidth 
can introduce systematic uncertainties in the fitting. 
However, since the abundances were subsequently
tied together in all annuli, we found that they were relatively
unaffected by this procedure (Sect~\ref{sect_bandwidth}).
We discuss the likely impact of uncertainties in 
the background  in Sect~\ref{sect_background}.

\subsection{The large-aperture spectrum} \label{sect_largeap}

\begin{deluxetable*}{llllll}
\tablecaption{Emission-weighted abundances within the single 3\arcmin\ (20~kpc) aperture\label{table_largeap}}
\tablehead{
\colhead{Model} & \colhead{\zfe} &
\colhead{\zo/\zfe} &
\colhead{\zne/\zfe} & \colhead{\zmg/\zfe} &\colhead {\zsi/\zfe} \\
}

\startdata
\multicolumn{6}{l}{0.5--7.0~keV}\\\hline
1T & 0.77$^{+1.66}_{-0.32}$ &$0.23^{+0.32}_{-0.23}$ & 1.21$^{+0.55}_{-0.41}$ & 
1.08$\pm0.31$ & 1.08$\pm0.59$ \\
2T & 1.09($>$0.60)& 0.02($<$0.20) & \nodata & \nodata & \nodata \\ \hline
\multicolumn{6}{l}{0.4--7.0~keV}\\\hline
1T & $1.10^{+2.18}_{-0.31}$ & $0.34^{+0.29}_{-0.21}$ & 1.21$^{+0.53}_{-0.41}$ &
1.04$\pm0.31$ & 1.02$\pm 0.57$\\
2T & $>$1.03 & $0.15\pm0.14$ & \nodata & \nodata & \nodata 
\enddata
\tablecomments{Abundance measurements and 90\% confidence intervals 
determined from fitting
the large-aperture spectrum. We present both single temperature (1T) and 
two-temperature (2T) results for the bandpass 0.5--7.0~keV and 0.4--7.0~keV.}
\end{deluxetable*}

We modelled the spectrum  with  a  hot gas component,
 a term to account for unresolved point-source emission and a component to 
accommodate the mild flaring. To model the hot gas, we adopted the APEC model 
of \citet{apec}, since it was the most up-to-date plasma code which is 
widely available (the impact of using a MEKAL model instead is 
addressed in Sect~\ref{sect_plasma}). 
To account for unresolved point-sources, we included a 
bremsstrahlung model with ${\rm kT= 7.3}$~keV, which was found to
give an excellent fit to the composite spectrum of the resolved sources
(Paper I). To model the flaring, we used the parameterization adopted 
in Sect~\ref{sect_flare} for the flaring on the S3 chip, albeit with
the normalization free. The components representing emission from the 
galaxy were additionally modified by photoelectric absorption due to 
cold material in the line-of-sight. The absorbing column was fixed at 
the Galactic value for the appropriate pointing 
\citep{dickey90}, and we adopted the absorption cross-sections of 
\citet[][but the results were unaffected if we adopted
those of \citealt{balucinska92}]{morrison83}.

We allowed \zfe\ to be free and initially tied the abundances of 
all remaining elements to Fe in their appropriate solar ratios. 
This model gave a good fit to the data ($\chi^2$/dof=121.04/114),
with \zfe$>1.5$. However, there were obvious sytematic discrepancies
between the model and the data at low energies (\ltsim 0.7~keV). 
We experimented with adding an additional hot gas component at a 
different temperature and, alternatively, systematically freeing 
individual metal abundances. In practice, the low-energy
residuals were only significantly reduced by freeing O, for which
the improvement in $\chi^2$ was  $>99.9$\% significant, on the 
basis of an F-test. We obtained best-fit abundances of
\zfe=0.77$^{+1.66}_{-0.32}$ and \zo/\zfe=$0.23^{+0.32}_{-0.23}$,
for a gas temperature of 0.57$\pm0.02$~keV. The fit was 
excellent ($\chi^2$/dof=106/113).
We experimented with systematically freeing the remaining abundances.
Although there was no significant improvement in the fit statistic following
this procedure, we  were able to estimate reasonable 90\% confidence limits
for the abundances of Ne, Mg, and Si
(Table~\ref{table_largeap})\footnote{Note: since the abundances ratios of 
Ne, Mg and Si to Fe were all 
consistent with unity, we determined the confidence region for each one 
of these elements assuming the abundance ratios for the 
others were {\em exactly} unity. This step was necessary to obtain interesting 
constraints}. The spectrum and best-fit model are shown in 
Fig~\ref{fig_spectra}. The lines of Ne, Mg and Si are clearly
visible in the spectrum, and so it is unsurprising that we have 
been able to constrain their abundances. 
{In order to minimize the number of variable parameters, we kept fixed the shapes of the flare and point-source model components. To determine the impact of this assumption on our error-bars, we additionally fitted the 3\arcmin\ aperture spectrum simultaneously with the flare and composite point-source spectra. Both the best-fit values and the error-bars were minimally affected by this procedure. We further investigate the impact of varying these components in Sect~\ref{sect_syserr}.}

Although the low-energy
residuals were substantially reduced by freeing \zo, there
was still some evidence of a feature at these energies (although, due 
to the good fit at higher energies, it was 
not reflected transparently in a poor
overall $\chi^2$ value). We found that adding an additional hot gas component 
(which we might expect given the temperature gradient observed in 
Sect~\ref{sect_annuli}) led to a statistically significant 
improvement in fit (at 99\% significance, on the basis of an F-test).
The quality of the fit was excellent ($\chi^2$/dof=97/111).
The temperatures of the two hot-gas components were found to be
0.33$^{+0.06}_{-0.04}$ and 0.72$^{+0.95}_{-0.15}$~keV (in good agreement
with the range of temperatures seen over the temperature gradient;
Sect~\ref{sect_annuli}). The best-fitting abundances are shown in
Table~\ref{table_largeap}, and clearly indicate $\sim$solar Fe abundance, 
and significantly  sub-solar O.
We experimented with allowing the remaining element abundances to vary,
as for the single-temperature case. However, we found that none of the 
abundances (excepting \zfe\ and \zo) could be constrained.
{Since adopting an over-simplistic spectral model can give rise to spurious abundances, we additionally tested the ``power law DEM'' (PLDEM) model of \citet{buote03a}, in which the differential emission measure is a power law
function of the temperature, and the power law exponent and range of 
gas temperatures are determined as fit parameters. The inferred abudances 
(\zfe$>$1.1) and (\zo/\zfe$<0.22$) were fully consistent with 
our two-temperature results. This agreement, which was also seen in 
NGC\th 5044 \citep{buote03b}, gives us confidence in the validity of the 
abundances derived from our two-temperature parameterization of the spectrum.
This is further supported by the good agreement with the spatially-resolved
analysis, below.}

The ability to  obtain  reliable abundance measurements is critically 
dependent on our continuum emission constraints. Especially when there
are multiple spectral components (\ie\ hot gas, unresolved point sources
and a flare component), it is essential to constrain the continuum at both
ends of the spectrum. Below $\sim$0.5~keV, the spectrum is relatively 
free from strong emission lines, and so the data are extremely useful to
this end. Unfortunately, the calibration of the \acis\ chips is more uncertain
in this regime, in part due to the build-up of carbanaceous deposits 
on the optical filter. Nevertheless, since we were using the most up-to-date 
correction for this quantum-efficiency degradation, we experimented with
extending the bandpass down to 0.4~keV. 
Using either 1 or 2 temperature best-fit model for the 0.5--7.0~keV bandpass,
we noticed a slight, systematic over-estimation of the count-rate below 0.5~keV
(the discrepancy being more pronounced for the 2 temperature case). 
Fitting the data in the extended band, 
we found systematically higher values of \zfe\ and \zo/\zfe\
for each model (Table~\ref{table_largeap}), but the remaining 
abundance ratios were relatively unaffected.  Although
there are calibration uncertainties in this regime, we observed very similar behaviour
using an AO1 observation of \srcthree\  (Sect~\ref{sect720}),
for which the \acis\ contamination was far less severe, which gave us confidence in
these results. 

\subsection{Spatially-resolved spectroscopy} \label{sect_annuli}
Observations of giant galaxies and groups often identify strong
temperature gradients. In order to investigate any such effects,
we performed spatially-resolved spectroscopy in a series of concentric 
annuli, centred on the peak of the X-ray emission. 
If there is a significant temperature gradient in the data,
on account of the quasi-ellipsoidal distribution of the hot gas in 
the galaxy halo,
the projection of the spectrum onto the sky does not accurately 
reflect the physical conditions at any point in the gas. In order to 
account for this effect, we employed 
the ``onion-peeling'' spectral deprojection algorithm
outlined in \citet{buote00c}. To determine the extent to which such
projection effects may alter our results, we additionally fitted 
traditional models directly to the spectra projected onto the sky.
Henceforth we distinguish between projected and deprojected results by
the soubriquets ``2D'' and ``3D'', respectively.

The quality of our data did not enable us to determine the metal
abundance independently in each annulus, preventing our 
investigation of any abundance gradients. For the 2D fitting, 
to maximize our constraints on the data, we fitted the projected
spectra simultaneously and tied equivalent metal abundances
between each annulus. 
Since the magnitude of flaring
did not vary strongly with position on the S3 chip (Sect~\ref{sect_flare}),
we were able to constrain the relative normalization of the flare component in 
each annulus to scale with the area of the extraction region,
allowing only its total normalization to be fit. Since the fraction
of emission arising from flaring differed in each annulus, this procedure
enabled us to disentangle its contribution more effectively than for our
large-aperture analysis (reflected in better constraints on our parameters).
Analogously to our large-aperture fitting, 
we initially allowed \zfe\ to be fitted freely and tied the remaining
elements to Fe in their appropriate solar ratios. 
To begin, we  tested the hypothesis of no temperature gradient 
(\ie\ tying the gas temperature between each annulus).
This isothermal model gave \zfe$>2.8$, but was statistically unacceptable 
($\chi^2$/dof=260.6/179). Freeing individual abundances did not enable
an acceptable fit to be found, allowing us to reject the hypothesis
of isothermality. 

We obtained a statistically significant improvement
in the fit  (at $>$99.99\% likelihood, on the basis of an f-test)
if we allowed the temperature in each bin to be determined
independently, although the fit was still unacceptable
($\chi^2$/dof=217.6/175). Examination of the spectra revealed significant 
residuals at low energy, especially in the innermost annulus, analogous
to those seen in the large aperture analysis. 
We therefore experimented with allowing the abundances of each metal in turn
to be determined independently from \zfe, although significant improvements in 
the fit were only seen when O was freed (which fell to $<0.15$). 
Adopting this model, we obtained a good fit to the data 
($\chi^2$/dof=189.4/174).
We experimented with adding extra hot gas components to the fit to 
determine whether they led to significant improvements. However, 
we found no evidence of a significant change in either the 
abundances or the fit statistic. 
The best-fitting value for \zfe\ (shown in Table~\ref{table_abundances})
was consistent with solar values (the lower limit at 99\% significance
on \zfe\ was 0.53). Furthermore, the gas was clearly 
deficient in O as compared to solar values; the 99\% upper-limit 
on the \zo/\zfe\ ratio was 0.24.

\begin{figure*}
\plottwo{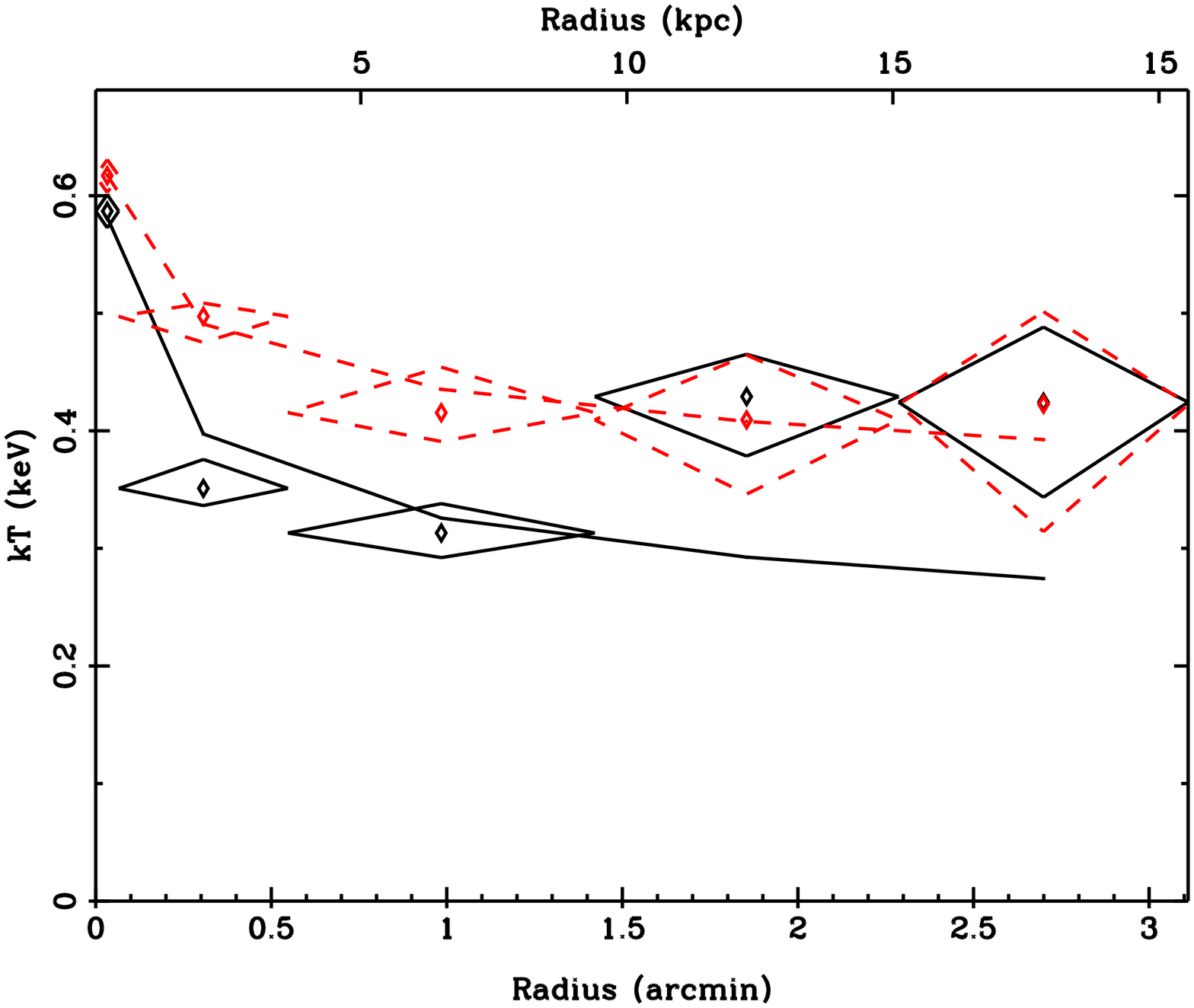}{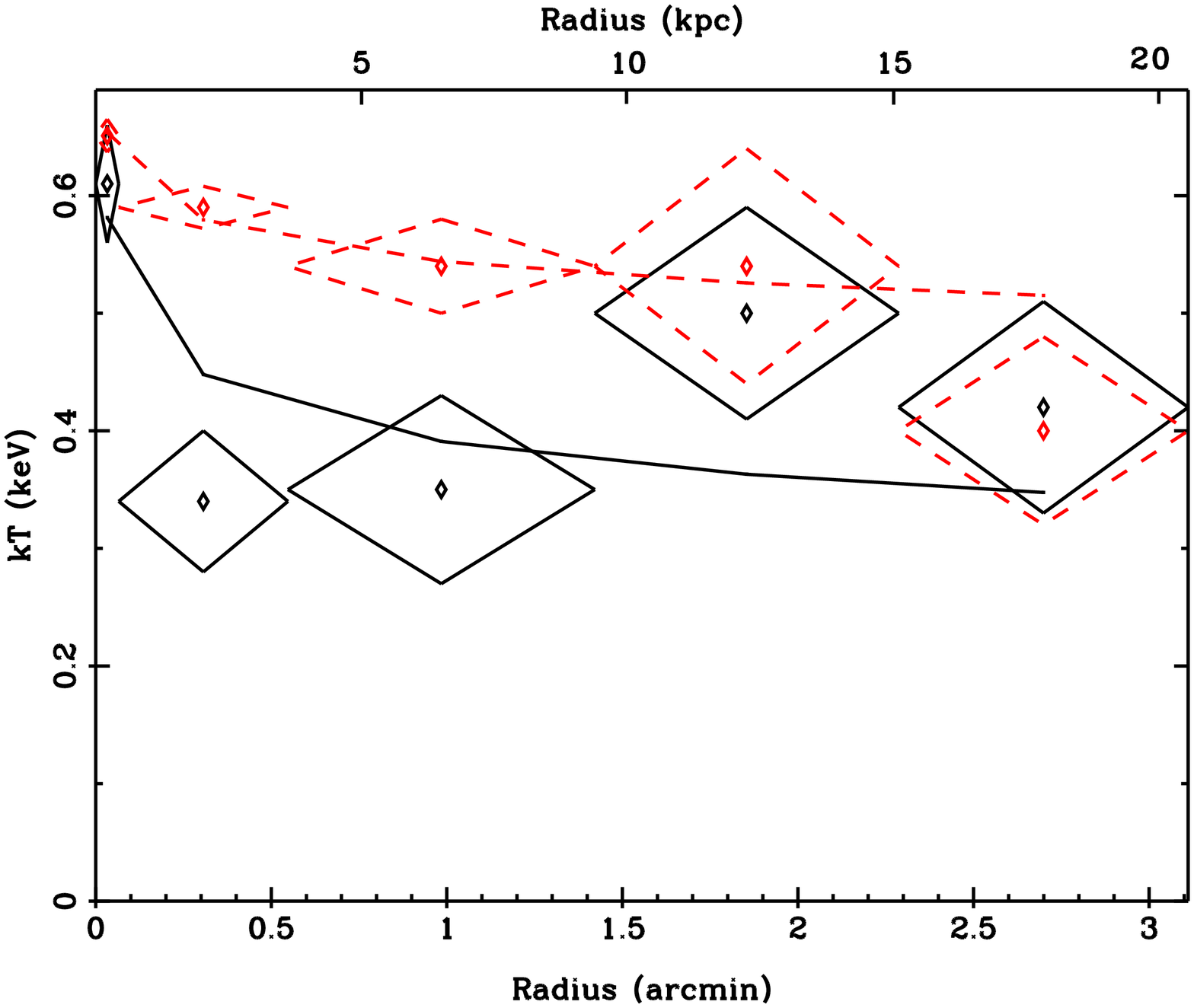}
\caption{Left panel: the temperature profile determined by fitting the
projected spectra. We show results determined both with all elements
tied to Fe
(dotted-lines) and with \zo\ free (solid line). In addition, we show 
a best-fit power law model to the profile. Right panel: the same results,
except fitting the deprojected spectra. For the deprojected analysis,
it was necessary to fix \zfe\ and \zo\ values at their best-fit values.} 
\label{fig_temp_profile}
\end{figure*}

The temperature profile was relatively isothermal, excluding a small peak in 
the centre  (Fig~\ref{fig_temp_profile}), although it was rather jagged in 
the second and third annuli. This is probably not a real effect; in fact
kT in the outer two annuli is expected to be systematically high due to the
truncation of the low-energy pass-band (Sect~\ref{sect_bandwidth}).
Furthermore,  due to the low temperature of the hot
gas, the important O lines are blended to some extent with the Fe L-shell 
``hump''. Since the shape of the ``hump'' strongly constrains the temperature,
this produces some parameter degeneracy between \zo\ and kT.
As there were possible systematic errors in our determination of \zo\ 
(Sect~\ref{sect_syserr}), we investigated the impact of such uncertainties
by measuring the temperature profile with all elements tied to Fe.
We found that the fit in this case was somewhat smoother, and the temperatures
systematically higher (Fig.~\ref{fig_temp_profile}), although there was
still evidence of a central temperature peak. 

As for the single-temperature, large-aperture analysis, we next 
systematically freed individually the remaining abundances and determined 
appropriate confidence regions for the parameters. Although we did not
find any evidence of an improvement in the fit statistic by this method,
this simply reflects best-fitting abundance ratios with respect to 
Fe close to solar (at least for those elements which could be constrained).
We found that we were able to constrain Si, Ne and Mg in addition to O
and Fe. In contrast to our large-aperture analysis, we were able to obtain
reliable confidence intervals when all of these metals were simultaneously
freed. The best-fit abundances are shown in Table~\ref{table_abundances}.
It is interesting to note that the abundance ratios were somewhat better constrained
than the absolute abundance of Fe. This is easily understood since the primary
source of uncertainty on the abundances is our constraint on the continuum
level.
If the continuum is over-estimated, the equivalent widths
of all the lines will be systematically under-estimated, and vice versa, producing
a strong correlation between the abundances (manifesting itself in long, thin
confidence regions in parameter space). 
We used a slightly modified implementation of the standard \xspec\ 
{\tt vapec} model, in which the absolute abundance of 
Fe was directly determined, but for the other elements the fit parameters
were the abundance {\em ratios} (in solar units) with respect to \zfe.
This enabled us to measure directly the errors on the abundance ratios in
\xspec\ without having explicitly to take account of this correlation. 
In previous studies \citep[\eg][]{buote03b}, we adopted Monte-Carlo 
simulations to achieve this.

For our 3D analysis, the inflated uncertainties in the profile 
introduced by the deprojection method were prohibitive if 
the metal abundances were to be determined by the fitting. We therefore
fixed the metal abundances to the best-fit values obtained from 2D
fitting and performed the deprojection using the 
algorithm outlined in \citet{buote00c}. To investigate the sensitivity
of the temperature profile to our abundance determination, we additionally
considered the case when \zo/\zfe$=1$. 
We found the results in both cases were very similar
to equivalent results found for our 2D analysis
(Fig~\ref{fig_temp_profile}). The temperatures
tended to be a little higher in each annulus using 3D fitting, 
although this was most pronounced for the \zo/\zfe=1 case, which cannot
acceptably fit the spectra all annuli for either 2D or 3D fitting.
In contrast, the deprojected model with free 
\zo\ was able to give good fits in all annuli. 
The spectrum of the innermost bin and the best-fitting 3D model is shown
in Fig~\ref{fig_spectra}.  Since the temperature 
profile was relatively flat (excepting the innermost bin), projection effects 
were not especially important. In the innermost bin, where there is a 
temperature gradient, due to the relatively
small spectral extraction region projection effects were also minor 
(Fig~\ref{fig_spectra}).

\begin{deluxetable*}{llllllllll}
\tablecaption{Abundance measurements and error budget\label{table_abundances}}

\tablehead{
\colhead{Par.} & \colhead{value} &
\colhead{${\rm \Delta}$Stat.} &
\colhead{${\rm \Delta}$calib} & \colhead{${\rm \Delta}$bkd} 
& \colhead{$\Delta$code} & \colhead{$\Delta$flare} & 
\colhead{$\Delta$bandw.}& \colhead{$\Delta$sources} & 
\colhead{$\Delta$\nh}\\}
\startdata
\zfe & 1.10 & $^{+1.7}_{-0.4}$ &  $\pm 0.3$  & $\pm 0.02$&$-0.16$& $\pm 0.02$& $\pm 0.33$& $-0.10$    & $\pm 0.01$\\
\zo/\zfe & 0.05 & $^{+0.10}_{-0.05}$&$+0.3$ & $\pm 0.01$& $-0.02$& $\pm 0.01$& $+0.11$   & $\pm 0.01$ & $\pm 0.01$\\
\zne/\zfe & 0.99 & $\pm 0.20$ &   $\pm 0.1$ & $\pm 0.03$& $-0.03$& $\pm 0.02$& $\pm 0.03$& $\pm 0.01$ & $\pm 0.02$\\
\zmg/\zfe & 1.07 & $\pm 0.27$ &   $\pm 0.08$& $\pm 0.02$& $+0.33$& $\pm 0.06$& $\pm 0.04$& $+0.03$    & $\pm 0.02$\\
\zsi/\zfe & 0.94 & $\pm 0.53$ &   $\pm 0.15$& $\pm 0.05$& $+0.18$& $\pm 0.03$& $\pm 0.06$& $+0.05$    & $\pm 0.02$
\enddata
\tablecomments{Emission-weighted abundance measurements for the
spatially-resolved spectroscopy, {fitting a single-temperature model for the hot gas as a function of radius}. In addition to the statistical 
(Stat.) errors, we present an estimate of the possible
magnitude of uncertainties on the abundances due to 
calibration (calib), background (bkd), plasma code (code),
flare model (flare), bandwidth (bandw.), unresolved source spectrum
(sources) and \nh. It is difficult to assess the impact of multiple 
sources of systematic error {\em simultaneously} affecting our results
(and it certainly is not correct simply to combine the errors in
quadrature), so it is probably most correct to assume that the source
of the largest systematic error dominates the systematic uncertainties. 
The magnitudes of the systematic errors listed are the changes in best-fit
values, which were often more sensitive to such errors
than the confidence regions (see text).
}
\end{deluxetable*}

\section{Systematic errors} \label{sect_syserr}
In order to estimate the magnitude of any systematic errors on our 
data-analyis,
we considered explicitly a number of effects, which are discussed in 
detail here. Those readers uninterested in the technical details can
proceed directly to Sect~\ref{sect720}. A summary of the magnitudes of
each systematic effect is given in Table~\ref{table_abundances}.
The discussion herein focuses primarily on our simultaneous 2D 
spatially-resolved spectroscopy results, since they gave us the best
 constraints.

\subsection{Calibration issues} \label{sect_calibration}
A number of aspects of the satellite calibration 
may have possible systematic effects on our 
best-fit results. Using the standard
\ciao\ data-processing software it was not possible to correct S3
chip data for ``charge transfer inefficiency'' (CTI), which degrades the spectral
resolution of the data. We were, however, able
to correct for a systematic drift in the instrumental gain by 
applying the {\tt apply\_gain}\footnote{http://cxc.harvard.edu/cont-soft/software/corr\_tgain.1.0.html}
task to our data  An alternative procedure which allows compensation for the 
effects of CTI was proposed by  
\citet{townsley02}\footnote{http://cxc.harvard.edu/cont-soft/software/ACISCtiCorrector.1.37.html},
although the calibration for this method is far less up-to-date.
Using the CTI-corrected data (without the gain-correction), 
we accumulated spectra in a set of concentric
annuli to match our spatially resolved spectroscopy.  We obtained a good fit 
($\chi^2$/dof=184.1/173) with the same spectral model, but with systematically 
lower overall Fe abundance (\zfe=$0.79^{+1.06}_{-0.26}$). Of the
abundance ratios, only \zo/\zfe\ was significantly altered (becoming
$0.33^{+0.23}_{-0.14}$). We obtained a very similar temperature profile
to that found for our standard analysis, although the central
peak was less pronounced (kT=$0.56^{+0.02}_{-0.08}$keV). 
It is difficult to disentangle to what extent 
the change in the best-fit abundances arose from the effects of the CTI 
correction, and to what extent they were simply a product of the less up-to-date
calibration of this technique. Nonetheless, these values give an
estimate of the calibration uncertainty due to the failure to correct
for CTI. 

It is also interesting to compare the results obtained for different
releases of the \chandra\ \caldb; processing the data using version 2.25,
for which the S3 spectral responses are known to have been subject to 
a small systematic error
led to systematically higher values of \zfe\ and \zo/\zfe\ 
($\Delta$\zfe$\simeq$0.2 and $\Delta$\zo/\zfe$\simeq$0.15). We can also
use this discrepancy as a rough estimate of the magnitude of any
possible calibration uncertainties remaining. 

\subsection{The background templates} \label{sect_background}
A potentially serious source of error in the determination of metal
abundances is an uncertainty in the background. 
Since the soft X-ray background is known to exhibit some dependence on
the pointing, there were possible systematic errors in our background
estimation at low energies (the background template was found to match
our observation well at high energies).
In order to determine the extent to which 
background errors may be affecting our results, we simply experimented
with altering the background normalization by $\pm$15\%, {comparable to
the 10\% variation between background fields found by 
M. Markevich\footnote{see http://cxc.harvard.edu/contrib/maxim/acisbg/COOKBOOK}}.
The best-fit metal abundances and temperature profiles showed no significant
changes due to  this procedure (Table~\ref{table_abundances}), 
and the confidence intervals were not appreciably affected. Our results, therefore, were 
relatively insensitive to errors in the  background estimation.

\subsection{The plasma codes} \label{sect_plasma}
In order to investigate the extent to which the choice of plasma code model
can influence our results, we experimented with replacing the APEC
hot gas model with a  MEKAL \citep{mekal1,mekal4} component, since the 
atomic physics, particularly of the Fe L-shell line, is somewhat different.
In general, both models
were able to fit the data, although MEKAL gave systematically poorer
$\chi^2$ values (for example, the best-fit $\chi^2$/dof for our 
best simultaneous fit to the spatially resolved spectra
was 202.2/174, as compared to 189.4/174 for the APEC model). 

We consistently obtained Fe abundances which were $\sim$10--20\% lower with the 
MEKAL code (and a similar effect with the \zo/\zfe\ value, although this
was poorly-defined due to the large error-bars).The best-fit 
MEKAL abundances for this fit were 
\zfe=0.94$^{+2.8}_{-0.42}$ and \zo/\zfe=$0.01^{+0.19}_{-0.01}$.
Freeing up other abundances, we found good agreement with the 
APEC results for Ne and Si. However, \zmg/\zfe\ was 
systematically higher ($1.43\pm0.40$), although marginally consistent
within error-bars with the  APEC result.
\citet{buote03b} noted a systematic
discrepancy of $\sim$10--20\% between the shapes of the APEC and MEKAL
spectral codes, in the vicinity of the Fe L-shell lines. This effect 
resulted in ${\rm Z_{Fe}^{apec}>Z_{Fe}^{mekal}}$ for single-temperature fits,
exactly as observed for \src. Since the Mg lines are close to the strong Fe 
L-shell lines
(although, at the temperature of the system, they do not overlap),
there is probably some interplay between them in the fit. 

\subsection{The flaring background model} \label{sect_flare_bg}
In order to obtain formally acceptable spectral-fitting results, it was 
necessary to include a term to account for the residual flaring, even 
though we filtered the data carefully to remove the strongest flares.
The flare was most evident above $\sim$2~keV and had 
uniform surface brightness across the interesting regions of the 
S3 chip. Since the spectral shape of the flare was not extremely 
well-constrained (due to the quality of the data) and since there were some
discrepancies in its spectrum when estimated by two different
means, this component was naturally a potential source of systematic 
errors. In order to determine the magnitude of any possible bias introduced
by incorrect flare modelling, we investigated the effect of replacing 
our preferred model with the alternative one based 
on our S1 chip flare estimate. This gave a slightly worse fit 
($\chi^2/dof$=193.7/174)
but there were no significant changes in the best-fit abundances,
although the error-bar on \zfe\ was slightly widened so that the 90\%
upper-limit became 3.5. 
The best-fitting temperatures in each annulus were also not significantly
changed by this procedure.

As a further test, we fitted the data by omitting completely any term
to account for flaring. In the outermost annuli (where the contribution due
to flaring is most significant due to the larger extraction area), the 
model significantly underestimated the data. Since the spectral shape of
the flare is significantly harder than that of the unresolved point-source
component, the unresolved emission term could not move to compensate 
entirely for the 
flaring contribution, so that the overall fit was formally unacceptable
($\chi^2/dof$=227/175). Nonetheless, neither the best-fitting abundance values nor 
the confidence ranges associated with them were dramatically altered
from our best-fit values ($\Delta$\zfe$\simeq\pm0.3$,  
$\Delta$\zo/\zfe$\simeq\pm0.01$). 
Examination of the residuals to the fit indicated that the model had simply 
fitted at low energies at the expense of the data \gtsim\ 2~keV.
The temperatures in all annuli were 
consistent, within error-bars, with the best-fitting 2D solution. However, 
the temperature was systematically lowered 
(by $\sim$0.05~keV and 0.1~keV) 
in the  outermost two annuli, respectively. 
Nonetheless, our abundance measurements
appear relatively insensitive to the modelling of the flare.

\subsection{The bandwidth} \label{sect_bandwidth}
The ability of the spectral-fitting models to determine reliable metal 
abundances and temperatures is dependent to some extent upon the 
bandwidth used in fitting \citep[see][]{buote00c}. 
In order to investigate the impact of narrowing our bandwidth, we first
experimented by truncating the data above 2.0~keV. Such a procedure
may be adopted to remove data heavily-biased by flaring, rather than
the modelling we used.
In adopting this procedure, we found that (omitting a 
flaring model component), we were able to obtain a good fit to the data
($\chi^2$=158.2/152). There were no significant changes in the 
best-fitting abundances or the temperature profile considering the size of the 
statistical error-bars.
As might be expected, however, this procedure dramatically 
increased our error bars on the absolute metal abundances; the
upper limit on \zfe\ was no longer constrained, although the 
lower limit was not  dramatically altered. 
The strong Fe L-shell
lines in the 0.7--1.2~keV region play a vital role in determining both 
the temperature and the Fe abundance of the gas, and so it is
unsurprising that we were still able to obtain a reliable gas temperature.
However, since we did not have the higher-energy information,
we were less able to constrain the continuum, hence the poorer
constraints on our parameters. 

In order to investigate the impact of our low-energy energy-cut, alternatively
we truncated the data below 0.7~keV and re-fitted the models. In this case,
we found a downward shift in the best-fit iron abundance,
but the predominant effect was to degrade significantly the error-bars 
on our results; we found that \zfe$\ge0.39$ and
\zo/\zfe$\le0.61$.
The strongest relevant O lines are found at $\sim$0.5--0.8~keV,
so it is hardly surprising that removing the data at low energies 
reduced our ability to constrain  the O abundance. The low-energy
data were also critical in enabling us to constrain the continuum level,
especially as there were multiple components being fitted in each annulus.
Failing to constrain the continuum translates into poorly-determined
metal abundances since it introduces serious uncertainties into 
the determination of line equivalent widths.
Furthermore, the
temperature profile was smoothed somewhat by this procedure, predominantly 
through an increase in the best-fitting temperature in the third annulus  
(kT$\sim$0.38~keV) and a slight widening of the error-bars. 
Since we were forced to exclude data below $\sim$0.7~keV in the outer two 
annuli during our analysis, this effect can explain the apparent
``jaggedness'' of the temperature profile in these data-bins. 
The best-fit abundances will not have been as strongly affected by our 
truncating the bandwidth in these annuli (as is confirmed by the agreement
between the two-temperature single aperture fit and the spatially-resolved
analysis), since the abundances were tied between all the annuli.  

Since the 
background began to dominate the data at energies $<$0.5--0.7~keV,
depending on the annuli, it was not possible to extend the bandwidth
downwards in our annuli spectra to determine the impact of this 
procedure. However, for our large aperture extraction,
it was possible to extend the bandwidth to 0.4~keV (although the 
calibration of the S3 chip is much more uncertain in this regime),
as discussed in Sect~\ref{sect_largeap}. This
increased the measured \zfe\ and \zo/\zfe\ values somewhat
(a similar effect was seen by \citealt{buote03b}), although the 
other abundance ratios were relatively unaffected.
\subsection{The unresolved source component}
We included a component to account for unresolved point-source emission in 
our data. The composite spectrum of all the resolved point sources in the
\dtwentyfive\ region of \src\ could be fitted very well by a single
bremsstrahlung model with temperature of 7.3~keV (see Paper I). Although
this is in excellent agreement with observations of other early-type
galaxies \citep{irwin03a}, the spectra of fainter LMXB (which would be 
unresolved) tend to be somewhat harder than those of brighter
objects \citep[\eg][]{church01}. This discrepancy is likely to be exacerbated
if high/ soft-state black-hole binaries are more common amongst the 
brightest source population of a galaxy. To determine the sensitivity
of our abundance measurements to the precise shape of this 
component, we next allowed the temperature of the bremsstrahlung
term  to fit freely. We found that its temperature tended to increase
(although, due to the limited band-width of our spectra, it 
could not be constrained). The resulting temperature profile and 
abundances were in good agreement, within errors, with our best-fitting
values.

\subsection{Column-density variation}
In our fitting, we fixed the neutral hydrogen column density to the
canonical Galactic value estimated for the particular pointing
\citep{dickey90}. In order to investigate any possible systematic 
effects arising from a mis-estimation of this value, we additionally
fitted the data but allowed the \nh\ to fit freely. The best-fitting column-density,
\nh${\rm =(2.9^{+9.3}_{-2.5})\times 10^{20}\ cm^{-2}}$,
was in excellent agreement with our adopted canonical value
(\nh${\rm =2.2\times 10^{20}\ cm^{-2}}$)
and we  did not find any significant improvement in the fit statistic 
($\Delta \chi^2$=0.1) by allowing \nh\ to be free. Since our best-fit
\nh\ was so close to the canonical value, we found that 
freeing \nh\ had no significant effect on the metal abundance
or temperature profile determination.

\section{The metallicity of \srcthree} \label{sect720}

\begin{figure}
\plotone{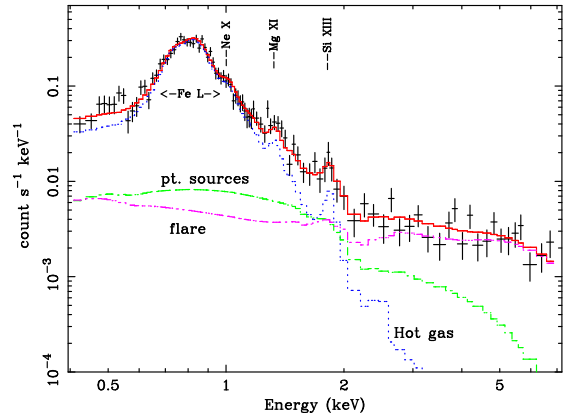}
\caption{The best-fitting spectrum to the innermost annulus 
(0--0.9\arcmin; 0--7.1~kpc) of the \srcthree\ data.
The data are shown, along with the best-fitting model.
We also show the spectrum decomposed into 
each component; the hot-gas (APEC) model is shown as a dotted line,
the unresolved point-source component (bremsstrahlung) is shown as 
a dot-dash line. The flare component is shown with a dash-dot-dot line.
} \label{fig720}
\end{figure}

\srcthree\ is a very isolated,  moderate-\lx\ elliptical galaxy with a 
slightly higher \lx/\lb\ than \src\ and significant hot gas emission. 
Since it was observed by \chandra\ during AO1, it provides an excellent
opportunity to investigate metal abundances in another moderate \lx/\lb\
system. 
\citet{buote02b} reported \chandra\ analysis results for
\srcthree, in which they were able to place tight constraints
upon the shape of the dark matter halo. Unfortunately,
there were strong periods of flaring in the background during the 
observation, heavily contaminating the spectrum of the diffuse gas
at energies \gtsim 2.0~keV. Since detailed spectral modelling was
not required in their analysis, these authors were able to mitigate
against the effects of flaring by truncating the spectrum above 2~keV.
With such a truncated energy-band, it was, however, not possible
to constrain the metal abundances (\cf\ Sect~\ref{sect_bandwidth}).
We re-analyzed the \chandra\ data of 
\srcthree, adopting the techniques described in Sect.~\ref{sect_flare}
to estimate the spectrum of the flare from the S3 chip. 
The data were reprocessed
identically to \src, and the periods of the strongest flaring were 
removed, giving a total exposure of 29.3~ks. 
The flare spectrum could be parameterized
adequately using the same model (broken power law plus broad
gaussian line) as used for \src. Although the fit was not formally
acceptable ($\chi^2$/dof=234.4/196), the residuals predominantly
appeared to be associated with low-amplitude, 
narrow features in the spectrum, so that the parameterization was a 
good approximation to the overall flare shape. Since the flaring
dominated the spectrum only over a small number of energy bins
(at the highest energy), such features were relatively unimportant, and 
our failure to account for them simply degraded the $\chi^2$ slightly.

We extracted spectra  from three annuli (with outer radii 0.9\arcmin,
1.7\arcmin\ and 2.4\arcmin) and centred at the X-ray centroid,
computed in the same manner as for \src. The annuli were chosen to 
contain approximately the same number of photons, and extended
out to 2.4\arcmin; we did not extend the annuli out further since the 
data were there very heavily contaminated by background flaring.
We fitted the data simultaneously in the range 0.5--7.0~keV, 
with a model comprising a term for the hot gas (APEC), a term
to account for unresolved point source emission (a 7.3~keV bremsstrahlung
component), and the flare model. 
We found there was little difference in the quality of the 
fit if we allowed the temperatures to be determined separately 
or if they were tied between each annulus, and so we tied the temperatures
to improve our abundance and temperature constraints. 
Similarly, we tied the metal abundances between
each annulus so as to determine an emission-weighted average. 
We initially tied all the metal abundances 
to \zfe, in their appropriate solar ratios, 
but we obtained a marginally unacceptable fit ($\chi^2$/dof=400.0/345).
We next experimented with allowing each metal in turn to fit
freely of Fe, and found a statistically significant improvement in the
fit only if the O was fitted freely.  
We obtained a good fit to the data ($\chi^2$/dof=368.2/344)
and the abundances were found to be 
\zfe=0.47${\rm ^{+0.36}_{-0.13}}$ and \zo/\zfe=${\rm 0.07^{+0.28}_{-0.07}}$
and kT=0.55$\pm0.02$~keV

Next, exactly as for the large aperture analysis of \src, we experimented
with extending the bandwidth down to 0.4~keV, as the count-rate was 
still significantly in excess of the background in each annulus at these energies. 
Since \srcthree\ was
observed during AO1, the degradation of the quantum efficiency at low
energies was a less significant effect and so the calibration of the S3 chip
in this range should be more secure. 
As in \src, we found that our best-fit to the 0.5--7.0~keV band  slightly overestimated 
the continuum between
0.4--0.5~keV (thereby leading to an under-estimate of the metal abundances). 
Fitting the model to the extended pass-band data, we obtained 
a good fit ($\chi^2$/dof=383.4/357)
and somewhat higher metal abundances, but consistent with the 
error-bars of the restricted energy-range solution. We determined
\zfe=$0.71^{+0.40}_{-0.21}$ and \zo/\zfe=$0.23\pm0.21$, and
kT=0.55$\pm0.02$. The 
lower-limit on \zfe, at the 99\% confidence limit, was \zfe=0.47. 
We experimented with freeing the remaining abundances, but were
only able to obtain constraints on Ne and Mg 
(\zne/\zfe=$0.44\pm0.43$, \zmg/\zfe=$1.26\pm0.35$). 
The strong flaring at 
high energies largely swamped the data in the vicinity of the Si line,
so that, in contrast to \src,  we could not obtain \zsi/\zfe\ constraints. 
The best-fitting spectrum for the innermost annulus is shown in 
Fig.~\ref{fig720}. These results were in general agreement with our 
results for \src\ (with the possible exception of  the \zne/\zfe\ ratio, 
which was $\sim$solar in \src, although the disagreement is marginal), 
and strongly
excluded very sub-solar metal abundances except for O, for which
\zo$\sim$0.2. 

\section{Discussion}
The \chandra\ observations of the normal, isolated, lenticular galaxy 
\src\ enabled us to place constraints on the metal abundances and 
the temperature profile of the hot gas. Revisiting the \chandra\ 
data of the isolated elliptical \srcthree\ we were also able to 
place good constraints on temperature and the metal abundances. 
The temperature profile of \src\ was relatively isothermal,
excepting a slight temperature peak in its centre. For \srcthree, we 
found that the data similarly showed evidence of  an $\sim$isothermal 
profile. In neither case did we see convincing evidence of a 
central temperature dip, such as is seen in ``cool core'' clusters and
groups. 
Both of these are normal, relatively isolated galaxies with moderate values of \lx/\lb, allowing us
to investigate the enrichment of early-type galaxies in a regime between
bright, gas-rich ellipticals (for which $\sim$solar abundances are
typically being found; see introduction) and the very gas-poor galaxies in 
which very low abundances are still being reported. 

\subsection{Near-solar Fe abundances}
In both of these galaxies, we were able to exclude the possiblity
of the highly sub-solar Fe abundances often reported for
early-type galaxies in the literature. These abundances
were frequently subject to the ``Fe bias'', on account of fitting a single
temperature model to intrinsically non-isothermal data \citep[\cf][]{buote99a}.
Multiple temperature components in a spectrum may arise from 
a strong temperature gradient over the spectral extraction region
or the mixing of photons due to projection effects (for 2D analysis). 
Alternatively, they may indicate a truly multiphase medium.
We note that the hard component due to
unresolved X-ray point sources in many early-type galaxies is too
dissimilar spectrally to the hot gas to induce the ``Fe bias''. 
{The presence of more than one spectral component in the single aperture spectrum of \src\ can clearly be understood in terms of the slight temperature gradient (which is not so large as to require deprojected analysis) evident in our spatially-resolved results. This is supported by the good agreement between our single-temperature spatially-resolved and two-temperature single aperture abundances. The ability of of a two-temperature model to parameterize a temperature gradient in the observing aperture, and still yield reliable abundances, is an established effect \citep[see \eg][]{buote03b}. 
It is possible that the spectrum is somewhat more complicated than this
picture. However, the general agreement between the PLDEM and the 
two-temperature fits in the single aperture, in addition to the agreement 
between the spatially-resolved and single aperture fit results, 
suggest that the
abundances are not especially sensitive to the precise parameterization,
provided it contains more than a single temperature component in this region. 
Similar conclusions were reached by \citet{buote03b}}.

{For the spatially-resolved spectroscopy of \src\ and \srcthree, the lack of a strong temperature gradient (which removed the need for deprojection) and the fact that the fits were not improved appreciably by using more than one hot gas component indicate that these data were}
{\em not} subject to the Fe bias. In fact, similar modelling of the \asca\ data
simply gave rise to very poor abundance constraints 
(\zfe$>0.3$ and \zfe$>0.2$, respectively:
\citealt{buote98c}, correcting to the solar abundances of \citealt{grsa}).
Interestingly, their best-fit  \asca\  values ($\sim$1.2 and $\sim$0.6) 
were in excellent agreement with our \chandra\ results.
Multi-temperature modelling of 
low-\lx\ systems with \asca\ or \rosat\ frequently 
gave similarly poor abundance constraints \citep[\eg][]{fabbiano94}, or were
subject to parameter degeneracy in which erroneously low 
values of \zfe\ could be obtained \citep[see][]{buote98c}.

Our \chandra\ results, and similar results in
the slightly lower-\lx/\lb\ radio galaxy NGC\th 1316 \citep{kim03b} now
reveal a self-consistent picture of $\sim$solar abundances from the 
centres of cool core clusters down to these moderate \lx\ galaxies. 
This is in contrast, however, to the extremely sub-solar abundances which
are still tending to be found in the lowest-\lx\ systems. If the low
signal-to-noise of the data does not bias these low abundances, the lack of 
enrichment in the lower-mass systems is a significant challenge to our 
understanding of the chemical enrichment process. Although the 
$\sim$solar abundances found in \src\ and \srcthree\ are far more 
consistent with more massive groups and clusters, 
we still did not obtain the several times solar 
values of \zfe\ typically predicted by enrichment models for individual
galaxies \citep[\eg][]{ciotti91,loewenstein91,brighenti99a}.

\subsection{Abundance ratio implications for supernova enrichment}
A number of effects can enrich or dilute the hot gas in an early-type
galaxy, such as supernovae, stellar mass-loss and gas inflow/ outflow
\citep[\eg][]{matteucci86}. The $\alpha$-elements are predominantly
processed in supernovae and so the abundance ratios of these metals with
respect to Fe provide a direct probe of the supernova enrichment 
history of the galaxy \citep[\eg][]{gastaldello02a}. 
Mass loss from intermediate-mass stars is expected to contribute significantly
to the ISM of an individual galaxy. However, 
assuming no nucleosynthesis of the $\alpha$-elements in the envelopes of 
intermediate-mass stars, the abundance ratios of the ejected gas 
simply reflect the ISM make-up at the time the star formed. For enrichment
dominated by very old stars, for example, stellar mass-loss will resemble
type II supernovae, which were primarily responsible for the early 
enrichment. 

In addition to Fe, we were able to obtain 
constraints for the abundance ratios of several other elements. 
Even taking into account the likely systematic effects observed in our data, 
the low \zo/\zfe\ ratio we measure in both galaxies seems a robust result.
Sub-solar values of \zo/\zfe\ have been reported, using a variety 
of instruments, in a variety of  other systems such as the 
bright group NGC\th 5044 \citep{buote03b,tamura03a}, M\th 87
\citep{gastaldello02a}, the massive elliptical NGC\th 4636
\citep{xu02a}, the starbursting galaxy M\th 82
\citep{tsuru97} and some clusters of galaxies 
\citep[][and references therein]{loewenstein04a}. 
We were also able to obtain constraints upon
Mg, Ne and (for \src), Si.
Intriguingly, we found that the  \zo/\zmg\ ratios were also significantly 
sub-solar. These elements
are primarily produced during Type II supernovae, for which models
tend to predict \zo/\zmg\ $\sim$0.9--1.9 
\citep[][converting to the solar abundances of \citealt{grsa}]{gibson97}.
Similar discrepancies have been found in several systems 
(\eg\ the centre of NGC\th 5044:\citealt{buote03b}; 
M\th 87: \citealt{gastaldello02a}). In addition, in \src, the \zo/\zne\
ratio is also dramatically lower than predicted from Type II enrichment
($\sim$1.3--1.9; \citealt{gibson97}), which exacerbates the discrepancy
in the \zo/\zmg\ ratios. In \srcthree, the \zo/\zne\ ratio is closer
to expectation, but the error-bars on the Ne abundance are large. 

Since low abundance ratios for O  have been
seen in a variety of galaxies with different instruments, it is unlikely to 
relate to calibration uncertainties or any bias due to an
abundance gradient (which we would not be able to constrain).
SNe metal yields are highly sensitive to theoretical modelling, 
so perhaps the simplest explanation for our results is a problem 
with the SNe models considered. \citet{loewenstein01} proposed that
a significant contribution of hypernovae to the enrichment of gas in 
rich clusters could explain similar low values of \zo/\zfe\
and \zo/\zsi, 
since the O-burning region is significantly enlarged in hypernovae. 
However, the hypernova models considered by 
\citet{umeda02a} did not reproduce very low values of both 
\zo/\zmg\ and \zo/\zne, as seen in our data,
although the authors acknowledge
that yields of Ne and Mg are sensitive to the assumptions of the model.
It may be of interest to note that recent \chandra\ observations
of the old SNR N49B revealed SN ejecta significantly enriched in  Mg, 
apparently without accompanying Ne or O enrichment \citep{park03}.

Notwithstanding these reservations, we were able to 
place some constraints upon the relative contribution of SN Type Ia
and Type II to the ISM enrichment.
In order to compare with previous results, 
we performed fitting analogous to that of \citet{gastaldello02a}.
Given the uncertainties in the hypernova yields, and to facilitate
a meaningful comparison, we did not include any 
enrichment contribution from hypernovae.
For a given fraction of enrichment arising from SNIa, f, 
theoretical metal yields from SNIa and SNII can be compared with
the data. We used the SNIa models of \citet{nomoto97b},
and the widely-adopted SNII yields taken from 
\citet[][hereafter N97]{nomoto97a}. In addition,
we also considered a range of SNII yields taken from 
\citet[][hereafter G97]{gibson97},
in order to examine the impact of the SNII model choice on our results.
It is worth noting that these models all assume a Salpeter IMF.
Although variation in the IMF is not expected to affect the abundance
ratios as significantly as the overall abundances 
\citep{gibson97}, it remains a further source of systematic uncertainty.

For the SNII yields of N97, we obtained $f=73\pm5$\%, which was relatively
insensitive to the SNIa model chosen (since, of the relevant elements,
only Si is significantly produced in SNIa, and the Si error bar is 
large). For the range of SNII models
considered in G97, we found systematically lower enrichment fractions,
reaching as low as $f=39\pm10$\%. For \srcthree, we found similar results
($f=85\pm 6$\% for the N97 model, and systematically
lower values if G97 yields were used, reaching as low as $f=62\pm12$\%).
Due to our large abundance ratio error bars, it was not possible to 
distinguish convincingly between different SNIa and SNII models.
One firm conclusion which we can draw, however, is that there is evidence of 
significant SNIa enrichment in both galaxies (f\gtsim0.4). Nevertheless,
for comparison with other work, we adopt the N97 results as a standard, 
but caution that systematic discrepancies still exist between SNII models,
which can significantly affect our results. It is interesting to compare our
results with  recent  determinations of f$\sim$80--90\% for the 
centre of  M\th 87, using the same yields \citep{gastaldello02a}. 
Similarly, in NGC\th 5044 \citet{buote03b} found  f$\sim$70--80\%,
and f$\sim$80--90\% was found in the centre of the bright nearby 
cluster A\th 1795 \citep{ettori02a}. The good agreement with our results
further confirms consistency in the enrichment mechanism
stretching from cluster-scales to moderate-\lx\ galaxies. 

Our fits to the data were, however, not formally acceptable 
for any mixture of SNIa and SNII model yields ($\chi^2$/dof$>4$).
Examination of the fit residuals revealed that, as expected, the low value 
of \zo/\zfe\ was problematic, even if we allowed the value for \src\ to 
increase by $\sim$0.3, consistent with systematic uncertainties in its value. 
Uncertainties in the hypernova yields of Ne and Mg \citep{umeda02a} 
make it difficult to investigate similarly any putative hypernova contribution,
which might account for the low \zo/\zfe\ ratio. If, instead, the 
low \zo/\zfe\ ratio reflects deficiencies in the SNII O yields,
it may be sufficient simply to omit this data-point from the fitting.
In this case we obtained good fits
(albeit for only 2--3 data points), with a systematically lower 
f for both galaxies. For \src, f=$55\pm7$\%, using N97; using G97 it can 
be as low as $16\pm13$\%. For \srcthree\ we found f=$68\pm8$\% for N97,
and systematically lower values down to $25\pm21$\% for G97.
In the centre of M\th 87 \citet{gastaldello02a} also found a formally 
unacceptable fit; recomputing f based on their data, but omitting \zo/\zfe,
we obtained a good fit with f=$77\pm5$\%. In the case of NGC\th 5044,
\citet{buote03b} computed f from the  Si and S ratios,
which were considered their most robust measurements, and so the low \zo/\zfe\
ratio did not similarly affect the results.
Therefore, even when O was excluded from the fitting, we found good
agreement between our results for \srcthree\ (and, to a lesser extent,
\src) and these more massive systems. In \src, the slightly lower 
value of f may reflect the increased significance of stellar-mass loss 
(from old stars) in a lower-mass system.

\begin{acknowledgements}
We would like to think Bill Mathews and Fabrizio Brighenti for helpful
discussions and carefully reading the manuscript. We would also like to 
thank  Fabio Gastaldello and Aaron Lewis for
helpful discussions concerning the data analysis and interpretation.
This research has made use of the NASA/IPAC Extragalactic Database (\ned)
which is operated by the Jet Propulsion Laboratory, California Institute of
Technology, under contract with the National Aeronautics and Space
Administration. Support
for this work was provided by NASA through \chandra\ award number
G02-3104X, issued by the Chandra X-ray Observatory Center, which
is operated by the Smithsonian Astrophysical Observatory for and
on behalf of NASA under contract NAS8-39073. Partial support for this
work was also provided by NASA under grant NAG5-13059, issued through
the Office of Space Science Astrophysics Data Program.
\end{acknowledgements}

\bibliographystyle{apj_hyper}
\bibliography{paper_bibliography.bib}

\end{document}